\documentclass[twocolumn,showpacs,aps,amsmath,amssymb.floatfix]{revtex4}

\begin{document}
\preprint{ND Atomic Theory 2001-4}

\title{Combined CI+MBPT calculations of energy levels
and transition amplitudes in Be, Mg, Ca, and Sr}

\author{I. M. Savukov}
 \email{isavukov@nd.edu}
 \homepage{http://www.nd.edu/~isavukov}
\author{W. R. Johnson}
 \email{johnson@nd.edu}
 \homepage{http://www.nd.edu/~johnson}
\affiliation{Department of Physics, 225 Nieuwland Science Hall\\
University of Notre Dame, Notre Dame, IN 46566}

\date{\today}

\begin{abstract}
Configuration interaction (CI) calculations in atoms with two valence
electrons,
carried out in the $V^{\rm (N-2)}$ Hartree-Fock potential of the core,
are corrected for core-valence interactions using many-body perturbation
theory (MBPT).
Two variants of the mixed CI+MBPT theory are described and applied to obtain
energy
levels and transition amplitudes for Be, Mg, Ca, and Sr.
\end{abstract}

\maketitle

\section{Introduction}

Although Be, Mg, Ca, and Sr atoms have been studied theoretically 
for many years and numerous calculations are available 
in the literature, energy levels of those divalent
atoms have been treated primarily with semiempirical methods and only a
limited number of low-lying levels have been evaluated using {\it ab-initio}
methods, which often do not provide 
sufficient precision or require extensive
computer resources.
Semiempirical methods, to their advantage, do not require significant
computer resources and can be applied easily to a large number of levels;
however, such theories have limited predictive power and accuracy.
Although energies obtained using semiempirical methods agree well
with one another and with experiment, oscillator strengths obtained by
different semiempirical calculations are inconsistent \cite{cimbpt:98}.
Examples of semiempirical calculations can be found for Be in \cite{cimbpt:98},
for Ca in  \cite{cimbpt:93}, and for Sr in \cite{cimbpt:95}.
Large-scale {\it ab-initio} configuration interaction (CI) calculations
of energies and transition rates, although capable of high accuracy,
have been performed only for a few low-lying levels in the
Be \cite{cimbpt:97a,cimbpt:01a} and Mg \cite{cimbpt:97b} isoelectronic
sequences.
The size of the configuration space in such CI calculations is limited by
the available computer resources.
Smaller-scale CI calculations, carried out in the frozen
$V^{\rm (N-2)}$ Hartree-Fock potential of the core, lead to poor results.
We found, for example, that frozen-core CI calculations in Ca
gave energies so inaccurate that it was difficult, if at all possible,
to identify many closely spaced levels of experimental interest.
Multi-configuration Dirac-Fock (MCDF) and Hartree-Fock (MCHF) methods
have also been used to obtain energies and oscillator strengths
in divalent atoms:
MCHF for Be-like ions \cite{cimbpt:97c}
and neutral calcium \cite{cimbpt:91},
and MCDF for Mg-like ions \cite{cimbpt:97d}. The accuracy
of MCHF and MCDF calculations
in neutral atoms is poor, basically because of computational limits
on the number of configurations.
Polarization potentials have been used in conjunction with
MCHF calculations \cite{cimbpt:93a} to
improve the accuracy of energies for Ca~I and Ca~II.
Many-body perturbation theory (MBPT)
calculations of energies and oscillator strengths for neutral divalent
atoms
using an effective Hamiltonian within a small model space,
are also found to be inaccurate \cite{cimbpt:97e,cimbpt:96b}.
Good agreement with experiment for divalent atoms, however, was
achieved
in Refs.~\cite{cimbpt:96a,cimbpt:01,cimbpt:98b} with a combined
CI+MBPT method. A related method
was applied to calculations of energies and oscillator strengths
for Mg-like ions in Ref.~\cite{cimbpt:98a}.
Among the {\it ab-initio} methods, CI+MBPT is particularly attractive since it
is capable of giving accurate energies and transition rates for both light and
heavy divalent atoms with modest computer resources.

A precise and efficient theoretical method for calculations of properties
of divalent atoms is needed for many possible applications of current interest, including
calculations of spectra, transition amplitudes, hyperfine structure
constants, polarizabilities, parity-nonconserving (PNC) amplitudes, van der
Waals coefficients, and Lennard-Jones coefficients. There is also growing
interest in properties of divalent atoms in conjunction with
low-temperature
Bose-Einstein condensation (BEC) experiments. For example, the prospect
for achieving BEC in divalent atoms was discussed in
\cite{cimbpt:99,cimbpt:00}
and depends on the size of the van der Waals coefficient.

At least two major difficulties have been recognized in studying divalent atoms.
First, core polarization effects are significant and must be taken into
account.
A similar situation exists in monovalent atoms where various methods
have been successfully applied to describe the
valence-core interaction. We have made extensive use of one of these
methods, MBPT, and have developed methods for calculating all diagrams
up to the third order for energies \cite{cimbpt:90}
and transition amplitudes \cite{cimbpt:00a}.
A second major
difficulty is that two valence electrons interact so strongly in neutral atoms
that
two-particle diagrams must be included to infinite order.
Since infinite order is required, the MBPT method is
difficult to apply. However, valence-valence correlations can be
accounted for completely using the CI method.

With this in mind, we have developed a method (similar to that used
in Refs.~\cite{cimbpt:96a,cimbpt:01,cimbpt:98b}
but with important differences)
for high-precision calculations of properties of
atoms with two valence electrons. The method starts with a complete
CI calculation of the interactions between the
valence electrons in a frozen core
and accounts for valence-core interactions using MBPT.
We apply this combined CI+MBPT method to calculate
energy levels and transition amplitudes for Be, Mg, Ca, and Sr.

\section{Method}

\subsection{Frozen-Core CI}

We start with a lowest-order description of a divalent atom in which the
closed N-2 electron core is described in the HF approximation and
valence or excited electrons satisfy HF equations in the ``frozen'' $V^{\rm
(N-2)}$
HF core.
As we mentioned in the introduction, the strong valence-valence correlations
must be
included to infinite order; the CI method accomplishes this.
The configuration space for divalent atoms is built up in terms of the
excited HF orbitals. We include all orbitals with angular
momentum $l \leq 5$ \ (partial wave contributions scale as
$1/(l+1/2)^{4}$) and we use 25 basis functions out of a complete set of 40
for each value of angular momentum.
The effect of these restrictions is insignificant considering
the perturbative treatment of valence-core correlations.

A detailed discussion of the CI method (as used here) can be found in
Ref.~\cite{cimbpt:93b}.
We introduce a configuration-state wave function $\Phi _{I}\equiv \Phi
_{JM}(ij)$
in which single-particle basis orbitals $i$ and $j$ are combined to give
a two-particle wave function with angular momentum $J$ and definite parity.
We then expand the general two-particle wave function $\Psi _{JM}$
in terms of all $\Phi _{JM}(ij)$ in our basis set
\begin{equation}
\Psi _{JM}=\sum\limits_{I}c_{I}\Phi _{I} .
\end{equation}
The expectation value of the Hamiltonian becomes
\begin{equation}
\left< \Psi_{JM} \left| H \right| \Psi _{JM} \right>
=\sum\limits_{I}E_{I}\, c_{I}^{2}+\sum\limits_{I,K}V_{IK}\, c_{I}\, c_{K} ,
\end{equation}
where $E_{I}=\epsilon _{i}+\epsilon _{j}$ is the sum of
single-particle HF energies and $V_{IK}$
is a first-order, two-particle correlation matrix element
(see, for example, \cite{cimbpt:93b})
between the configurations $I=(ij)$ and $K=(kl)$.
The variational condition leads to CI equations
\begin{equation}
\sum\limits_{K}\left( E_{I}\delta _{IK}+V_{IK}\right) c_{K}=\lambda\, c_{I} ,
\label{evp}
\end{equation}
from which CI energies ($\lambda$) and wave functions ($\sum_I c_I \Phi_I$)
are found.

\subsection{Combining CI with MBPT}

Core polarization effects can be treated using MBPT.
In this paper, we introduce two procedures that enable us to combine
frozen-core CI and second-order two-valence-electron MBPT,
which we refer to as
``CI averaging'' and ``Brueckner-Orbital CI'' methods.

\subsubsection{CI averaging}

In this first method, the core-valence interaction $\Delta E_{vc}$ is
obtained by ``averaging'' MBPT corrections over CI wave functions:
\begin{equation}
\Delta E _{vc}=\sum c_{I}\, c_{K}\left< \Phi _{I} \left| H^{(2)} \right| \Phi
_{K}\right> \, ,
\end{equation}
where the configuration weights $c_{I}$ and $c_{K}$ are taken from
the solution of the CI equation, Eq.~(\ref{evp}), and $H^{(2)}$ is that part
of the effective Hamiltonian
projected onto the valence electron subspace containing second-order
valence-core interactions.
The dominant second-order parts of the effective Hamiltonian, beyond those
accounted for in the CI calculation,
are the
screening and self-energy diagrams: $H^{(2)}= H^{\rm screen}+ H^{\rm self}$,
the self-energy being much larger than the screening and both being larger
than the remaining second-order terms.

We borrow ready-to-use formulas, derived using standard
techniques, from Ref.~\cite{cimbpt:96b}.
The screening contribution to the effective Hamiltonian is
\begin{align}
H^{\rm screen}_{v^{\prime}w^{\prime }vw} =& -\eta _{v^{\prime }w^{\prime }}
\eta_{vw} \sum\limits_{{ \alpha ^{\prime }\beta ^{\prime }
\alpha \beta }}C_{1}(\alpha ^{\prime }\beta ^{\prime }\alpha \beta) \times
 \nonumber \\
&\sum\limits_{nbk}\frac{(-1)^{j_{w^{\prime }}+j_{v}+j_{n}+j_{b}}}{[k]}
\left\{
\begin{array}{ccc}
j_{\alpha ^{\prime }} & j_{\beta ^{\prime }} & J \\
j_{\beta } & j_{\alpha } & k
\end{array}
\right\} \times \nonumber\\[0.5ex]
&\hspace{3em} \frac{Z_{k}(\alpha ^{\prime } b\alpha n)Z_{k}(\beta ^{\prime}n\beta
b)}
{\epsilon _{\beta }+\epsilon _{b}-\epsilon _{\beta ^{\prime}}-\epsilon _{n}}\
,
\end{align}
where 
\begin{gather}
C_{1}(\alpha ^{\prime }\beta ^{\prime }\alpha \beta
)=(-1)^{J}\left[ \delta _{\alpha ^{\prime }v^{\prime }}\delta
_{\beta ^{\prime }w^{\prime }}\delta _{\alpha v}\delta _{\beta
w}+\delta _{\alpha ^{\prime }w^{\prime }}\delta _{\beta ^{\prime
}v^{\prime }}\delta _{\alpha w}\delta _{\beta
v}\right]\hspace{0em}\nonumber\\
 +\, \delta _{\alpha ^{\prime }v^{\prime
}}\delta _{\beta ^{\prime }w^{\prime }}\delta _{\alpha w}\delta
_{\beta v}+\delta _{\alpha ^{\prime }w^{\prime }}\delta _{\beta
^{\prime }v^{\prime }}\delta _{\alpha v}\delta _{\beta w}  .
\end{gather}
The self-energy contribution to $H^{(2)}$ is
\begin{gather}
H^{\rm self}_{v^{\prime }w^{\prime }vw}=\eta _{v^{\prime }w^{\prime }}\eta
_{vw}
\left[ \delta _{w^{\prime }w}\Sigma _{v^{\prime }v}+\delta _{v^{\prime
}v}\Sigma _{w^{\prime }w} \right. \nonumber\\
\left. +\ (-1)^{J}(\delta _{v^{\prime }w}\Sigma _{w^{\prime
}v}+\delta _{w^{\prime }v}\Sigma _{v^{\prime }w})\right] \, ,
\end{gather}
where
\begin{gather}
\Sigma _{ij}(\epsilon _{0})=\sum\limits_{kcmn}
\frac{
(-1)^{j_{m}+j_{n}-j_{i}-j_{c}}}{[j_{i}][k]}\frac{X_{k}(icmn)Z_{k}(mnjc)}
{\epsilon _{0}+\epsilon _{c}-\epsilon _{m}-\epsilon _{n}} \hspace{0em}
\nonumber\\
+\ \sum\limits_{kbcn}
\frac{(-1)^{j_{i}+j_{n}-j_{b}-j_{c}}}{[j_{i}][k]}\frac{X_{k}(inbc)Z_{k}(bcjn)}
{\epsilon _{0}+\epsilon _{n}-\epsilon _{b}-\epsilon _{c}} \ .
\label{self-energy}
\end{gather}
In the above equations, $J$ is the angular momentum of the coupled 
two-particle states. The coupled radial integrals $X_k(abcd)$ and
$Z_k(abcd)$ are defined in \cite{cimbpt:96b}. We use the notation
$[k] = 2k+1$. The quantities $\eta_{vw}$ are normalization constants,
$\eta_{vw} =1/\sqrt{2}$ for identical particle states and 1, otherwise.
In the expression for the self-energy, the angular momenta of the 
$i$th and $j$th orbitals satisfy $\kappa_{i}=\kappa _{j}$, where
$\kappa_i=\mp (j_i+1/2)$ for $j_i=l_i \pm 1/2$ is the angular quantum
number uniquely specifying the spinor for state $i$.
Since we found that the second-order self-energy correction
is very important, we also consider the fourth-order self-energy
obtained by iteration:
\begin{equation}
\Sigma_{ij}(\epsilon_0) \rightarrow \Sigma_{ij}(\epsilon_0) +
\sum_{k\neq i}
\frac{\Sigma_{ik}(\epsilon_0)\Sigma_{kj}(\epsilon_0)}{\epsilon_i-\epsilon_k} \ .
\end{equation}
In heavy atoms, the choice of $\epsilon _{0}$ deserves
special consideration.
Problems with denominators arise from the fact that single-particle
orbitals used in the self-energy calculation are not optimal, in the sense
that there is mutual interaction between valence electrons not accounted for,
even approximately, in the $V^{(N-2)}$ potential and accounted for
excessively in the $V^{(N)}$ potential which is used,
for example, in Ref.~\cite{cimbpt:01}. One practical solution to this problem
is to use ``optimized'' denominators \cite{cimbpt:01}. A consistent
theory requires an {\it ab-initio} treatment of the denominator problem.
Basing calculations of atoms with two valence electrons on a more realistic
potential can reduce uncertainties in the choice of the denominator in
the self-energy corrections.

We calculated energies of
several levels using the CI averaging method and found that the best agreement
with experiment for Be and Mg was obtained with $\epsilon _{0}$ equal to
1/2 of the CI energy. For the
case of Ca, the best agreement was obtained choosing $\epsilon _{0}$ between
1/2 and 1 times the CI energy. One advantage of the CI averaging method is
that the basic CI code is simple and that the CI wave functions can
be stored and used many times. A cut-off condition can be imposed,
as a compromise between speed and accuracy. The fastest approximation (giving
the poorest accuracy) is obtained by restricting the MBPT corrections to the
leading
configurations. We used this leading configuration approximation to estimate
the magnitude of the core-excitation effects as the first step in developing
our
computer code. Adjusting the cut-off condition, we readily reached a high
level of accuracy (finally we chose the cut-off condition $|c_{I}c_{K}|<0.002$
for all calculations). The energies for several states of Be, Mg, and Ca
presented in this paper have been calculated with the CI averaging method.
The principal drawback of this method is that wave functions necessary
for calculations of other properties are not automatically obtained.

\subsubsection{Brueckner-Orbital CI}

The effective Hamiltonian formalism \cite{cimbpt:96b} leads to the problem of
diagonalizing
the Hamiltonian matrix built on the frozen-core two-electron configuration
state functions $\Phi _{I}$. We split this matrix into
functionally distinct pieces:
\begin{equation}
H = H^{(0)}+H^{(1)}+H^{(2)} ,
\end{equation}
where $H^{(0)}$ is the zeroth-order Dirac-Fock Hamiltonian, which in the 
Dirac-Hartree-Fock
(DHF)
basis is
\[
H^{(0)}_{v^\prime w^\prime vw} =
\delta _{vv^\prime} \delta _{ww^\prime}
\left( \epsilon_{v}^{0}+\epsilon _{w}^{0}\right) ,
\]
and $H^{(1)}$ is the first-order electron-electron interaction Hamiltonian
\[
H^{(1)}_{v^\prime w^\prime vw}=V_{v^\prime w^\prime vw}^{(1)} ,
\]
defined in Ref.\cite{cimbpt:96b}.
$H^{(2)}$ is the second-order correction which consists of the two-particle
screening correction and the one-particle self-energy correction
defined previously.
In the CI averaging method, $H^{(0)}+H^{(1)}$ is diagonalized first
in a DHF basis (where $H^{(0)}$ is diagonal)
to give state energies and CI wave functions,
then $H^{(2)}$ is evaluated using the CI wave functions to give corrections
for the core-valence interaction.

In the Brueckner-orbital (BO) CI method, the basis functions are chosen
as orthonormal solutions of the quasi-particle equation,
\begin{equation}
[h_{0}+ V_{\rm HF} + \Sigma_{ij}(\epsilon )] \phi_{j}=\epsilon^{BO} \phi_{i} .
\label{BOeq}
\end{equation}
In this BO basis, \begin{equation}
\left( H^{(0)}+H^{\rm self}\right)_{v^\prime w^\prime vw}
= \delta _{vv^\prime}\delta_{ww\prime}(\epsilon _{v}^{BO}+\epsilon _{w}^{BO})
.
\end{equation}
The basis orbitals include second-order self-energy corrections
together with the lowest-order DHF potential.
The residual non-trivial part of the effective Hamiltonian in the BO basis
is the sum $H^{(1)}+H^{\rm screen}$. In the Bruckner orbital-CI method, the
residual Hamiltonian matrix is evaluated in
the BO basis and diagonalized to obtain state energies and CI wave functions.
The BO-CI method is equivalent to CI averaging method if we
neglect energy differences in the denominators of $H^{\rm self}$
and $H^{\rm screen}$ (of order of the valence-valence interaction energy),
which  are small compared to the core excitation energies.
The BO-CI method is also equivalent to the effective Hamiltonian
method in \cite{cimbpt:01} to the same level of precision, provided all second-order
diagrams are included. Some advantage is gained in accuracy compared to
the CI averaging method, since the largest valence-core
corrections [those from $\Sigma _{ij}(\epsilon_0)$] are taken into account to
infinite order.

The Brueckner-orbital CI method is very convenient for calculations
of transition amplitudes; once the residual interaction is diagonalized,
the associated wave functions are
immediately available.
We include random-phase approximation (RPA) corrections
in calculations of transition amplitudes
by replacing ``bare'' matrix elements with ``dressed'' elements as explained
in \cite{cimbpt:00a}.
Length-form and velocity-form dipole matrix elements are found
to be in close agreement in BO-CI calculations that include RPA corrections.

\section{Calculations of spectra using CI averaging}

The CI averaging method is fast and convenient for calculations of energies
when a large number of levels are needed, especially at the stage of
adjusting the code parameters. Below, we present
our calculations for many levels of Be, Mg, and Ca atoms to demonstrate
the accuracy of this method.
We evaluate the valence-core correction $\Delta E_{vc}$ to the CI energy using
a subset
of the CI coefficients limited to those satisfying $|c_I\, c_K| \leq 0.002$.
The parameter $\epsilon_0$ in the self-energy was chosen to be $\epsilon_{\rm
CI}/2$ for Be and Mg. For calcium it was increased to $3\epsilon_{\rm CI}/4$
to obtain
better agreement for energies of the $4p^2$ states.

The basis set used to set up the calculations consisted of 25/40 DHF basis
functions for each value of $l\leq 5$. The basis functions were formed as
linear combinations of B-splines of order 7, constrained to a cavity of
radius $R$=80 $a_0$.

\subsection{Calculations for Be}

\begin{table}
\caption{Comparison of CI-averaging energy levels (cm$^{-1}$)
of Be I with experimental data from the NIST database \protect\cite{nist:01}.
\label{Table1}}

\begin{ruledtabular}
\begin{tabular}{lllcc}
Config. & Term & J & NIST & CI-average \\ \hline
2s$^{2}$ & 1S & 0 & 0 & 0 \\
2s2p & 3P$^{o}$ & 0 & 21978 & 21996 \\
2s2p & 3P$^{o}$ & 2 & 21981 & 22000 \\
2s3s & 3S & 1 & 52081 & 52074 \\
2p$^{2}$ & 1D & 2 & 56882 & 56890 \\
2s3p & 3P$^{o}$ & 1 & 58907 & 58890 \\
2s3p & 3P$^{o}$ & 2 & 58908 & 58896 \\
2p$^{2}$ & 3P & 1 & 59695 & 59749 \\
2p$^{2}$ & 3P & 2 & 59697 & 59747 \\
2s3d & 3D & 3 & 62054 & 62033 \\
2s3d & 1D & 2 & 64428 & 64414 \\
2s4s & 3S & 1 & 64506 & 64528 \\
2s4s & 1S & 0 & 65245 & 65261 \\
2s4p & 3P$^{o}$ & 2 & 66812 & 66792 \\
2s4d & 3D & 3 & 67942 & 67924 \\
2s4f & 3F$^{o}$ & 3 & 68241 & 68224 \\
2s4f & 1F$^{o}$ & 3 & 68241 & 68224 \\
2s4d & 1D & 2 & 68781 & 68774 \\
2s5s & 3S & 1 & 69010 & 69056
\end{tabular}
\end{ruledtabular}

\end{table}
We chose to study a Be atom for several reasons. First, this atom has a small
core and, consequently, requires relatively little computation time.
Second, because of the small size of the core-valence interaction,
calculations for Be are expected to be very precise.

A comparison of the resulting CI energies with measured energies
from the National Institute of Standards and Technology (NIST) database \cite{nist:01}
is shown in Table~\ref{Table1}. This comparison provides the first test of
the CI averaging method.
The values listed in the table agree with experiment at the level of
tens of cm$^{-1}$. The residual deviation can be explained as neglect of
small Coulomb and Breit diagrams, which will be the subject of future
investigations.

It is also interesting to compare CI energies, with and without the MBPT
corrections $\Delta E_{vc}$, with energies from the NIST database. Such a
comparison is given in
Table~\ref{Tabe3} and illustrates the importance of the valence-core
corrections.
\begin{table}
\caption{Comparison of frozen-core CI energies (cm$^{-1}$) and CI-averaging
energies for Be I with experimental energies from the NIST database \protect\cite{nist:01}.
}\label{Tabe3}

\begin{ruledtabular}
\begin{tabular}{lllrcrcr}
\multicolumn{1}{c}{Config}&
\multicolumn{1}{c}{Term} &
\multicolumn{1}{c}{J} & \multicolumn{1}{c}{NIST} &
\multicolumn{1}{c}{CI-average}&
\multicolumn{1}{c}{Diff.} &
\multicolumn{1}{c}{Frozen CI}&
\multicolumn{1}{c}{Diff.} \\
\hline
2s3s & 1S & 0 & 54677 & 54664 & -13 & 54509 & 168 \\
2p$^{2}$ & 3P & 0 & 59694 & 59737 & 43 & 60090 & -396 \\
2s5s & 1S & 0 & 69322 & 69307 & -15 & 69387 & +65
\end{tabular}
\end{ruledtabular}
\end{table}

The agreement with experiment improves by an order of magnitude for the
CI-averaging method as compared with a frozen-core CI calculation.
Indeed, we found it necessary to use the more precise energies obtained
from the CI-averaging method to properly identify the transitions shown in
this table.

\subsection{Calculations for Mg}

Another example where the CI averaging method predicts energy levels
accurately is magnesium. In this atom, however, core correlations are larger
and the treatment of the valence-core interaction term requires more careful
analysis. One important aspect is
choosing the parameter $\epsilon _{0}$ in the denominators of
the MBPT corrections, another is the treatment of self-energy diagrams. We
found mild sensitivity of final energies in Mg to the choice of $\epsilon_0$.
The corrected energies shown in the column headed `CI + 2nd' in
Table~\ref{Tamg}, which were obtained with the choice
$\epsilon_0 = \epsilon_{\rm CI}/2$, are seen to be in
close agreement with experimental energies \cite{nist:01}.

\begin{table}
\caption{Comparison of energies (a.u.)\ in Mg obtained from frozen-core CI,
CI-averaging with 2nd-order self-energy, and CI-averaging with chained 4th-order
self-energy, with experimental energies from the NIST database
\protect\cite{nist:01}.
\label{Tamg}}
\begin{ruledtabular}
\begin{tabular}{lccccc}
Conf. Level & CI & CI+2nd & CI+4th & Expt. & $\Delta$ (cm$^{-1}$) \\
\hline
$3s^{2}$ \ \ $^{1}S_{0}$ & 0.818 & 0.8329 & 0.833513 & 0.833518 & 1 \\
$3s4s$ \ $^{1}S_{0}$ & 0.624 & 0.6349 & 0.635260 & 0.635303 & 9 \\
$3s5s$ \ $^{1}S_{0}$ & 0.583 & 0.5938 & 0.594240 & 0.594056 & 40 \\
$3s6s$ \ $^{1}S_{0}$ & 0.566 & 0.5772 & 0.577813 & 0.577513 & 66 \\
$3p^{2}$ \ \ $^{3}P_{0}$ & 0.562 & 0.5695 & 0.569747 & 0.570105 & 79 \\
$3s3p$ \ $^{3}P_{1}$ & 0.723 & 0.7336 & 0.733991 & 0.733869 & 27 \\
$3s3p$ \ $^{1}P_{1}$ & 0.661 & 0.6733 & 0.673673 & 0.673813 & 31 \\
$3s4p$ \ $^{3}P_{1}$ & 0.604 & 0.6156 & 0.615834 & 0.651524 & 68 \\
$3s4p$ \ $^{1}P_{1}$ & 0.597 & 0.6086 & 0.608606 & 0.608679 & 16 \\
$3s3p$ \ $^{3}P_{2}$ & 0.723 & 0.7333 & 0.733867 & 0.733684 & 67
\end{tabular}
\end{ruledtabular}
\end{table}

Typically, the self-energy correction is much
larger than other valence-core diagrams; for example, in the Mg ground
state, the self-energy is $-1.65\times 10^{-2}$ a.u.\ while the
screening contribution is ten times smaller, $1.83\times 10^{-3}$ a.u.
Valence-core contributions in fourth-order, obtained by iterating (or
chaining)
the second-order Brueckner corrections are also found to be significant,
$-6.57\times 10^{-4}$ a.u.\ for the Mg ground state. The effect of including
corrections from chaining the self-energy 
shown in the column headed `CI + 4th' in Table~\ref{Tamg} is seen to further
improve the agreement with experiment.

\subsection{ Ca atom}

\begin{table}[b]
\caption{Comparison of the accuracy of
frozen-core CI and CI averaging calculations for Ca.
The parameter $\epsilon_{0}=0.75\, \epsilon_{CI}$.}
\label{Taca1}
\begin{ruledtabular}
\begin{tabular}{llccccc}

Conf.& Level& frozen CI & Diff. & CI-average& Diff. & Expt. \\
\hline
4s5s & $^{1}S_{0}$ & 31901 & -1416 & 33196 & -121 & 33317 \\
4p$^{2}$ & $^{3}P_{0}$ & 36699 & -1718 & 38900 & 483 & 38418 \\
4s6s & $^{1}S_{0}$ & 39376 & -1314 & 40504 & -186 & 40690 \\
4p$^{2}$ & $^{1}S_{0}$ & 41480 & -306 & 42366 & 580 & 41786 \\
4s7s & $^{1}S_{0}$ & 42673 & -1604 & 43841 & -436 & 44277 \\
4s8s & $^{1}S_{0}$ & 44277 & -1610 & 45551 & -336 & 45887 \\
4s9s & $^{1}S_{0}$ & 45629 & -1206 & 46912 & 77 & 46835
\end{tabular}
\end{ruledtabular}
\end{table}

In Table~\ref{Taca1}, several even parity $J=0$ levels are calculated with the
frozen-core
CI and CI-averaging methods. Compared to the frozen-core CI method, the
agreement is significantly improved
with
the addition of MBPT corrections, changing the difference between experiment
and theory 
from approximately one thousand
cm$^{-1}$ to a few hundred cm$^{-1}$. This significant change
clearly indicates the importance of the valence-core interaction, which is
much stronger
than in the case of Be and Mg. As a result, the final accuracy of CI+MBPT
method is
also lower than for the lighter atoms. While the poor accuracy of frozen
CI energies prevents the identification of energy levels, more accurate
CI+MBPT energies
permit one to identify many Ca levels. It is
interesting to notice that the sequence of
experimental levels  for the states of
a particular symmetry is the same as the sequence of theoretical eigenvalues.
Once the question of classification is solved,
various properties of atoms can be calculated using, for example, frozen-core
CI.

In the case of Ca, another problem that needs attention is the
choice of
the parameter $\epsilon _{0}$ in the self-energy, the dominant part of
the core-valence interaction. We find that there is an optimal value of this
parameter between $\epsilon_{CI}/2$, our standard value for Be and Mg, and
$\epsilon_{CI}$, for which the ground state becomes very accurate.
In Table~\ref{Taca1} we chose this parameter to be $0.75\, \epsilon_{CI}$.
In the following section, we
will illustrate our calculations of transition amplitudes for 
several levels of Mg, Ca, and Sr where other precise calculations 
and measurements exist.

\section{Calculations using the Brueckner-orbital CI method}

In this section, we present our calculations of energies and transition
amplitudes with the Brueckner-orbital CI method. Our basis consisted of 25
$V_{HF}^{N-2}$ orbitals (those orbitals were constructed of 40 B-splines in
the cavity 80 a.u.), in which 14 lowest excited states were replaced with
Brueckner orbitals. The resulting one-valence electron energies for the
divalent atoms were tested
by comparing with experimental energies for the corresponding monovalent ions. 
For Mg$^{+}$, the BO energies agree with experiment better than do the
second-order energies (Table~\ref{Tamgbo}). 
\begin{table}[t]
\caption{Comparison of  DHF spline energies `DHF', second-order
energies `2nd order', and
energies resulting from diagonalization of the self-energy matrix,
Brueckner-orbital `BO' energies,
with experiment for the Mg$^{+}$ ion. The core configuration is $1s^22s^{2}2p^{6}$.
The size of the self-energy matrix is 14$\times$14 for each angular momentum. All
energies are expressed in cm$^{-1}$. }
\label{Tamgbo}
\begin{ruledtabular}
\begin{tabular}{ccccc}
States & DHF & 2nd order & BO & Expt. \\
\hline
3s$_{1/2}$ & 118825 & 121127 & 121184 & 121268 \\
4s$_{1/2}$ & 50858 & 51439 & 51446 & 51463 \\
5s$_{1/2}$ & 28233 & 28467 & 28469 & 28477 \\
3p$_{1/2}$ & 84295 & 85508 & 85542 & 85598 \\
4p$_{1/2}$ & 40250 & 40625 & 40633 & 40648 \\
5p$_{1/2}$ & 23642 & 23808 & 23811 & 23812
\end{tabular}
\end{ruledtabular}
\end{table}
A second iteration of the BO equation
was also included in the CI-averaging method (Table~\ref{Tamg}) to improve
accuracy. The small size of the residual deviation 
from experiment in both tables can be
attributed to higher-order diagrams.
Two-particle screening corrections with the restriction $n<15$ were included in
the effective
Hamiltonian, diagonalization of which provided the initial and final state
wave functions necessary for the calculation of transition amplitudes. We
checked that restrictions on the number of BO and screening
diagrams  included in the calculation did not lead to significant errors.
Dressed transition amplitudes were used to
take into account RPA corrections, which provide
better length- and velocity-form agreement. We completely neglected
the extremely
time consuming structural radiation
corrections which are expected to be small for the length form; for this
reason,
the result calculated in length form should be considered as more accurate.
Small normalization corrections are also omitted.

\subsection{Be case}
\begin{table}[tbp]
\caption{Comparison of the present transition energies $\omega$ (a.u.) and oscillator
strengths $f$ for Be with those from
other theories and experiment. A few allowed singlet--singlet transitions of
the type $S_{0}-P_{1}^{o}$ between low-lying states are considered. The
experimental uncertainties are given in parentheses.
}
\label{Tabebo}
\begin{ruledtabular}
\begin{tabular}{lllll}
\multicolumn{1}{c}{Transition} &
\multicolumn{1}{c}{Source} &
\multicolumn{1}{c}{$\omega$(Theory)} &
\multicolumn{1}{c}{$\omega$(Expt.)} &
\multicolumn{1}{c}{$f$} \\
\hline
2s$^{2}$-2s2p & present & 0.194126 & 0.193954 & 1.3750 \\
& \cite{cimbpt:72} & & & 1.38(0.12) \\
& \cite{cimbpt:74} & & & 1.34(0.05) \\
& \cite{cimbpt:98} & 0.91412 & & 1.375 \\
& \cite{cimbpt:93c}& 0.193914 & & 1.374 \\
& \cite{cimbpt:85} & & & 1.3847 \\
& \cite{cimbpt:81} & & & 1.470 \\
& \cite{cimbpt:96c}& & & 1.375 \\
\hline
2s$^{2}$-2s3p & present & 0.274231 & 0.274251 & 0.00904 \\
& \cite{cimbpt:98} & 0.27441 & & 0.00901 \\
& \cite{cimbpt:93c}& 0.274236 & & 0.00914 \\
& \cite{cimbpt:85} & & & 0.0104 \\
& \cite{cimbpt:81} & & & 0.037 \\
& \cite{cimbpt:97c}& & & 0.00885 \\
\hline
2s3s-2s2p & present & 0.054977 & 0.05519 & 0.1188 \\
& \cite{cimbpt:98} & 0.05509 & & 0.118 \\
& \cite{cimbpt:93c} & 0.055198 & & 0.1175 \\
& \cite{cimbpt:85} & & & 0.1199 \\
& \cite{cimbpt:81} & & & 0.140 \\
\hline
2s3s-2s3p & present & 0.025128 & 0.025107 & 0.9557 \\
& \cite{cimbpt:72} & 0.0252 & & 0.958 \\
& \cite{cimbpt:93c} & 0.025124 & & 0.9565 \\
& \cite{cimbpt:85} & & & 0.9615
\end{tabular}
\end{ruledtabular}
\end{table}
The most accurate results for divalent atoms are expected for Be since it
contains the smallest MBPT corrections. In Table~\ref{Tabebo}, we compare our
calculations with available precise calculations and experiment. Transition
energies agree with experiment to better than 0.1\%, except for the transition
$2s3s ^{1}S-2s2p^{1}P$ which has 0.4\% accuracy.
Our oscillator strengths agree well with those obtained in very accurate
{\it ab-initio} calculations of Ref.~\cite{cimbpt:93c} and in semiempirical
calculations of Ref.~\cite{cimbpt:98} that
reproduce energies very closely; for the principal
transition $2s2 ^{1}S-2s2p^{1}P$, our value 1.375 differs by 1 in the 4th
digit from the value 1.374 in Ref.~\cite{cimbpt:93c},
the accuracy being better than 0.1\%, and coincides with the value of
Ref.~\cite{cimbpt:98}. Very close agreement with {\it ab-initio} theory is
also achieved for the transition $2s3s^{1}S-2s3p^{1}P$.
For suppressed transitions,
an accuracy of 1\% is obtained. Conducting
a simple statistical analysis, we found that energy differences in the
CI-averaging and BO-CI calculations have similar statistical errors, but
slightly different systematic shifts which can be explained partially by
different denominators in the two methods. Another reason is the cut-off
condition 0.002 in the former method and restriction on the number of 
Brueckner orbitals in
the latter. The effect of the partial wave restriction on the ground state
energy in both methods is 6 cm$^{-1}$.
If this value is accounted for, the agreement becomes slightly better.
The results in our tables are not extrapolated owing to the smallness of
the omitted partial wave contributions.
\subsection{The cases of Mg, Ca, and Sr}

The accuracy of both the CI-averaging and the BO-CI calculations considered
above decreases from light to heavy divalent atoms.
\begin{table}[t]
\caption{Comparison of BO-CI energies (cm$^{-1}$) with experiment for Mg, Ca, and Sr.
\label{Taboen}}

\begin{ruledtabular}
\begin{tabular}{cccc}
Levels & Theory & Expt. & Diff. \\
\hline
\multicolumn{4}{c}{Mg atom}\\[0.0pc]
3s4s $^{1}S_{0}$ & 43452 & 43503 & -51 \\
3s5s $^{1}S_{0}$ & 52517 & 52556 & -39 \\
3s6s $^{1}S_{0}$ & 56154 & 56187 & -33 \\
3s3p $^{3}P_{1}$ & 21834 & 21870 & -44 \\
3s3p $^{1}P_{1}$ & 35059 & 35051 & 8\\
3s4p $^{3}P_{1}$ & 47806 & 47844 & -38 \\
3s4p $^{1}P_{1}$ & 49317 & 49347 & -30\\[0.0pc]
\multicolumn{4}{c}{Ca atom} \\[0.0pc]
4s5s $^{1}S_{0}$ & 33505 & 33317 & 188 \\
4p$^{2}$ \ $^{3}P_{0}$ & 38651 & 38418 & 233 \\
4s6s $^{1}S_{0}$ & 40862 & 40690 & 172 \\
4s4p $^{3}P_{1}$ & 15595 & 15210 & 385 \\
4s4p $^{1}P_{1}$ & 23797 & 23652 & 145 \\
4s5p $^{3}P_{1}$ & 36760 & 36555 & 205 \\
4s5p $^{1}P_{1}$ & 36917 & 36732 & 185 \\[0.0pc]
\multicolumn{4}{c}{Sr atom} \\[0.0pc]
5s6s $^{1}S_{0}$ & 30874 & 30592 & 282 \\
5p$^{2}$ \ $^{3}P_{0}$ & 35913 & 35193 & 720 \\
5p$^{2}$ \ $^{1}P_{0}$ & 37696 & 37160 & 536 \\
5s5p $^{3}P_{1}$ & 15081 & 14504 & 577 \\
5s5p $^{1}P_{1}$ & 21981 & 21699 & 282 \\
5s6p $^{3}P_{1}$ & 34293 & 33868 & 425 \\
5s6p $^{1}P_{1}$ & 34512 & 34098 & 414
\end{tabular}
\end{ruledtabular}
\end{table}
Table~\ref{Taboen} illustrates this tendency in BO-CI calculations: for Mg, the
theory-experiment differences range within 50 cm$^{-1}$, similar to what we
have in Table~\ref{Tamg}, and for Ca the deviation from experiment increases
to about 200 cm$^{-1}$ which is comparable to that in Table~\ref{Taca1}.
The lowest
accuracy is for Sr, which has the largest core and MBPT corrections. Similar
results for energies have been obtained in Ref.\cite{cimbpt:01}. Our
experiment-theory differences exhibit a systematic shift, which if
subtracted, brings results into better agreement. For example, in Ca this shift
is
216 cm$^{-1}$. After its subtraction, the residual deviation is
73 cm$^{-1}$. This subtraction procedure can be used in cases where closely
spaced levels are difficult to identify. The systematic shift can be
attributed to omitted correlations that affect mostly the ground state
which is used as a reference.
The cut-off condition in the CI-averaging method and restrictions on the
number of BO and screening diagrams also has some effect on the accuracy of
our results. This is one reason why the two methods give slightly different
energies. In future development of our computer code, we will try to remove
such restrictions completely. Another reason why the two methods give different 
results is that the choices of
$\epsilon_{0}$ were different.
In Table~\ref{mgcasr}, we illustrate our calculations of transition amplitudes
for Mg, Ca, Sr. All of our transition amplitudes completely agree with those of
recent CI+MBPT calculations by \citet{cimbpt:01},
and are close to experimental values. Length-form and velocity-form amplitudes
agree to better than 1\% for allowed
transitions. 
\begin{table}
\caption{Comparison of our length-form (L) and velocity-form (V) 
calculations with those from
Ref.~\protect\cite{cimbpt:01}
and with experiment.
\label{mgcasr}}
\begin{ruledtabular}
\begin{tabular}{llll}
&
\multicolumn{1}{c}{Mg} &
 \multicolumn{1}{c}{Ca} &
\multicolumn{1}{c}{Sr} \\
\hline
\multicolumn{4}{c}{$^{1}P_{1}^{o}(nsnp)-^{1}S_{0}(ns^{2})$}\\
 L & 4.026 & 4.892 & 5.238 \\
 V & 4.019 & 4.851 & 5.212 \\
 Other\footnotemark[1] &
 4.03(2) & 4.91(7) & 5.28(9) \\
 Expt. & 4.15(10)\footnotemark[2]
& 4.967(9)\footnotemark[5]
& 5.57(6)\footnotemark[6]
\\
 & 4.06(10)\footnotemark[3]
 &4.99(4)\footnotemark[6]
 &5.40(8)\footnotemark[8]
\\
& 4.12(6)\footnotemark[4]
& 4.93(11)\footnotemark[7]
& \\[1ex]
\multicolumn{4}{c}{$^{3}P_{1}^{o}(nsnp)-^{1}S_{0}(ns^{2})$}\\
 L & 0.0063 & 0.0323 & 0.164\\
 V & 0.0070 & 0.0334 & 0.166 \\
 Other\footnotemark[1]
& 0.0064(7) & 0.034(4) & 0.160(15) \\
 Expt. & 0.0053(3)\footnotemark[9]
& 0.0357(4)\footnotemark[12]
& 0.1555(16)\footnotemark[15] \\
 & 0.0056(4)\footnotemark[10]
& 0.0352(10)\footnotemark[13]
& 0.1510(18)\footnotemark[13] \\
 & 0.0061(10)\footnotemark[11]
& 0.0357(16)\footnotemark[14]
& 0.1486(17)\footnotemark[16]%
\end{tabular}
\end{ruledtabular}
\footnotetext[1]{\citet{cimbpt:01}.}
\footnotetext[2]{\citet{cimbpt:80}.}
\footnotetext[3]{\citet{cimbpt:73}.}
\footnotetext[4]{\citet{cimbpt:66}.}
\footnotetext[5]{\citet{cimbpt:00}.}
\footnotetext[6]{\citet{cimbpt:80a}.}
\footnotetext[7]{\citet{cimbpt:83}.}
\footnotetext[8]{\citet{cimbpt:76}.}
\footnotetext[9]{\citet{cimbpt:92}.}
\footnotetext[10]{\citet{cimbpt:82}.}
\footnotetext[11]{\citet{cimbpt:75}.}
\footnotetext[12]{\citet{cimbpt:86}.}
\footnotetext[13]{\citet{cimbpt:97f}.}
\footnotetext[14]{\citet{cimbpt:80b}.}
\footnotetext[15]{\citet{cimbpt:84}.}
\footnotetext[16]{\citet{cimbpt:88}.}
\end{table}
Forbidden transitions are more problematic, owing to cancellation
effects, and have poorer agreement between gauges and with experiment. The
inclusion of the Breit interaction and negative-energy contributions, which
are more important for the velocity form, might improve the situation.
We also noticed that,
if the balance between states such as
$p_{1/2}$ and $p_{3/2}$ in relativistic basis is not properly maintained, the
results for nonrelativistically forbidden transitions will be unstable. In
addition, those transitions were affected by the number of BO and screening
diagrams included in calculations. To minimize or exclude those effects in
the BO-CI method, the BO orbitals and cut-off
conditions were made completely symmetric with respect to $l+1/2$ and $l-1/2$
orbitals and included BO and screening corrections with number of excited
orbitals less than 15.

\section{Summary and conclusion}
In this paper, we have introduced two methods to
improve the accuracy of the
frozen-core CI calculations using MBPT:
the CI-averaging method and the Brueckner-orbital CI
method. We have applied these methods to Be, Mg, Ca, and Sr atoms. Our
calculated energies and transition amplitudes for those atoms are
in close agreement with the results of the best available theories and
experiments. Compared to semiempirical theories, our method has an advantage in
accuracy, and compared to other {\it ab-initio} theories, an advantage of simplicity.
These two methods can also be used to evaluate properties of
Rydberg states for which only
semiempirical calculations exist. Further improvement in accuracy
is possible and is being pursued. This theory can be extended easily to treat
particle-hole excited states of closed-shell atoms, atoms with three
valence electrons, and other more complicated systems.

\begin{acknowledgments}
The authors are grateful to U. I. Safronova for helping to establish the correctness
of computer codes. We are thankful to M. Kozlov for discussion of the theory.
We thank H. G. Berry for reading manuscript and giving useful comments.
We thank A. Derevianko for pointing out about experimental interest in divalent
atoms. This work was supported in part by
National Science Foundation Grant No.\ PHY-99-70666.
\end{acknowledgments}

\bibliography{cmb2}
\end{document}